\documentclass[a4paper,twoside,twocolumn,english,aps,prc,superscriptaddress,floatfix]{revtex4}
\usepackage[T1]{fontenc}
\usepackage[latin1]{inputenc}
\usepackage{graphicx}
\usepackage{amssymb}

\makeatletter

 \newenvironment{lyxlist}[1]
   {\begin{list}{}
     {\settowidth{\labelwidth}{#1}
      \setlength{\leftmargin}{\labelwidth}
      \addtolength{\leftmargin}{\labelsep}
      }}
   {\end{list}}

\newcommand{\amev}[0]{$A$ MeV{}}

\usepackage{babel}
\makeatother
\begin{document}
%% $Id: paper.tex,v 1.2 2004/09/06 13:39:24 franklan Exp $

\title{Model-independent tracking of criticality signals in nuclear multifragmentation data} 

%%\input{authors.tex}
%% $Id: paper.tex,v 1.2 2004/09/06 13:39:24 franklan Exp $
%% List of authors for INDRA collaboration
%%
\author{J.D.~Frankland}
\affiliation{ GANIL, CEA et IN2P3-CNRS, B.P.~55027, F-14076 Caen Cedex, France.}
\author{A.~Chbihi}
\affiliation{ GANIL, CEA et IN2P3-CNRS, B.P.~55027, F-14076 Caen Cedex, France.}
\author{A.~Mignon}
\affiliation{ GANIL, CEA et IN2P3-CNRS, B.P.~55027, F-14076 Caen Cedex, France.}
\author{M.L.~Begemann-Blaich}
\affiliation{ Gesellschaft für Schwerionenforschung mbH, D-64291 Darmstadt, Germany.}
\author{R.~Bittiger}
\affiliation{ Gesellschaft für Schwerionenforschung mbH, D-64291 Darmstadt, Germany.}
\author{B.~Borderie}
\affiliation{ Institut de Physique Nucl\'eaire, IN2P3-CNRS, F-91406 Orsay Cedex,  France.}
\author{R.~Bougault}
\affiliation{ LPC, IN2P3-CNRS, ENSICAEN et Université, F-14050 Caen Cedex, France.}
\author{J.-L.~Charvet}
\affiliation{ DAPNIA/SPhN, CEA/Saclay, F-91191 Gif sur Yvette Cedex, France.}
\author{D.~Cussol}
\affiliation{ LPC, IN2P3-CNRS, ENSICAEN et Université, F-14050 Caen Cedex, France.}
\author{R.~Dayras}
\affiliation{ DAPNIA/SPhN, CEA/Saclay, F-91191 Gif sur Yvette Cedex, France.}
\author{D.~Durand}
\affiliation{ LPC, IN2P3-CNRS, ENSICAEN et Université, F-14050 Caen Cedex, France.}
\author{C.~Escano-Rodriguez}
\affiliation{ GANIL, CEA et IN2P3-CNRS, B.P.~55027, F-14076 Caen Cedex, France.}
\author{E.~Galichet}
\affiliation{ Institut de Physique Nucl\'eaire, IN2P3-CNRS, F-91406 Orsay Cedex, France.}
\affiliation{ Conservatoire National des Arts et Métiers, F-75141 Paris Cedex 03.}
\author{D.~Guinet}
\affiliation{ Institut de Physique Nucléaire, IN2P3-CNRS et Universit\'e F-69622 Villeurbanne, France.}
\author{P.~Lautesse}
\affiliation{ Institut de Physique Nucléaire, IN2P3-CNRS et Universit\'e F-69622 Villeurbanne, France.}
\author{A.~Le Fèvre}
\affiliation{ Gesellschaft für Schwerionenforschung mbH, D-64291 Darmstadt, Germany.}
\author{R.~Legrain}
\thanks{deceased}
\affiliation{ DAPNIA/SPhN, CEA/Saclay, F-91191 Gif sur Yvette Cedex, France.}
\author{N.~Le~Neindre}
\affiliation{  Institut de Physique Nucl\'eaire, IN2P3-CNRS, F-91406 Orsay Cedex, France.}
\author{O.~Lopez}
\affiliation{ LPC, IN2P3-CNRS, ENSICAEN et Université, F-14050 Caen Cedex, France.}
\author{J.~\L ukasik}
\affiliation{ Gesellschaft für Schwerionenforschung mbH, D-64291 Darmstadt, Germany.}
\affiliation{ Institute of Nuclear Physics, Pl-31342 Krak\'ow, Poland.}
\author{U.~Lynen}
\affiliation{ Gesellschaft für Schwerionenforschung mbH, D-64291 Darmstadt, Germany.}
\author{L.~Manduci}
\affiliation{ LPC, IN2P3-CNRS, ENSICAEN et Université, F-14050 Caen Cedex, France.}
\author{J.~Marie}
\affiliation{ LPC, IN2P3-CNRS, ENSICAEN et Université, F-14050 Caen Cedex, France.}
\author{W.F.J.~Müller}
\affiliation{ Gesellschaft für Schwerionenforschung mbH, D-64291 Darmstadt, Germany.}
\author{L.~Nalpas}
\affiliation{ DAPNIA/SPhN, CEA/Saclay, F-91191 Gif sur Yvette Cedex, France.}
\author{H.~Orth}
\affiliation{ Gesellschaft für Schwerionenforschung mbH, D-64291 Darmstadt, Germany.}
\author{M.~P\^arlog}
\affiliation{ National Institute for Physics and Nuclear Engineering, RO-76900 Bucharest-M\u{a}gurele, Romania.}
\author{M.~Pichon}
\affiliation{ LPC, IN2P3-CNRS, ENSICAEN et Université, F-14050 Caen Cedex, France.}
\author{M.~F.~Rivet} 
\affiliation{ Institut de Physique Nucl\'eaire, IN2P3-CNRS, F-91406 Orsay Cedex,  France.} 
\author{E.~Rosato}
\affiliation{ Dipartimento di Scienze Fisiche e Sezione INFN, Università di Napoli ``Federico II'', I-80126 Napoli, Italy.}
\author{R.~Roy}
\affiliation{ Laboratoire de Physique Nucléaire, Université Laval, Québec, Canada.}
\author{A.~Saija}
\affiliation{ Gesellschaft für Schwerionenforschung mbH, D-64291 Darmstadt, Germany.}
\author{C.~Schwarz}
\affiliation{ Gesellschaft für Schwerionenforschung mbH, D-64291 Darmstadt, Germany.}
\author{C.~Sfienti}
\affiliation{ Gesellschaft für Schwerionenforschung mbH, D-64291 Darmstadt, Germany.}
\author{B.~Tamain}
\affiliation{ LPC, IN2P3-CNRS, ENSICAEN et Université, F-14050 Caen Cedex, France.}
\author{W.~Trautmann}
\affiliation{ Gesellschaft für Schwerionenforschung mbH, D-64291 Darmstadt, Germany.}
\author{A.~Trzcinski}
\affiliation{ Soltan Institute for Nuclear Studies, Pl-00681 Warsaw, Poland.}
\author{K.~Turz\'o}
\affiliation{ Gesellschaft für Schwerionenforschung mbH, D-64291 Darmstadt, Germany.}
\author{A.~Van Lauwe}
\affiliation{ LPC, IN2P3-CNRS, ENSICAEN et Université, F-14050 Caen Cedex, France.}
\author{E.~Vient}
\affiliation{ LPC, IN2P3-CNRS, ENSICAEN et Université, F-14050 Caen Cedex, France.}
\author{M.~Vigilante}
\affiliation{ Dipartimento di Scienze Fisiche e Sezione INFN, Università di Napoli ``Federico II'', I-80126 Napoli, Italy.}
\author{C.~Volant}
\affiliation{ DAPNIA/SPhN, CEA/Saclay, F-91191 Gif sur Yvette Cedex, France.}
\author{J.P.~Wieleczko}
\affiliation{ GANIL, CEA et IN2P3-CNRS, B.P.~55027, F-14076 Caen Cedex, France.}
\author{B.~Zwieglinski}
\affiliation{ Soltan Institute for Nuclear Studies, Pl-00681 Warsaw, Poland.}
\collaboration{INDRA and ALADIN collaborations}
\noaffiliation

\begin{abstract}
We look for signals of criticality in multifragment production in
heavy-ion collisions using model-independent universal fluctuations
theory. The phenomenon is studied as a function of system size, bombarding
energy, and impact parameter in a wide range of INDRA data. For very
central collisions ($b/b_{max}<0.1$) we find evidence that the largest
fragment in each event, $Z_{max}$, plays the role of an order parameter,
defining two different regimes at low and high incident energy, respectively,
according to the scaling properties of its fluctuations. Data for
a wide range of system masses and incident energies collapse on to
an approximately universal scaling function in each regime for the
most central collisions. The forms of the scaling functions for the
two regimes are established, and their dependence on the total mass
and the bombarding energy is mapped out. Data suggest that these regimes
are linked to the disappearance of heavy residues in central collisions.
\end{abstract}

\date{$Revision: 1.2 $ $Date: 2004/09/06 13:39:24 $}

\maketitle

\section{Introduction\label{sec:introduction}}

It has long been hoped that the study of nuclear multifragmentation
reactions as observed in intermediate energy heavy-ion collisions
\cite{DeSouza91:ar+au_35-110amev,Bowman91:xe+au_50amev,Ogi91:rise_and_fall,Tsang93:onset_of_vaporization,Peaslee94:rise_and_fall_kr+au}
can give valuable information on the nuclear matter phase diagram
and equation of state \cite{Ber83:spinodal,Fai:1985tc,Peilert:1989kr}.
The principal guide-wire for this research has been the search for
signs of something analogous to a liquid-gas phase transition in data
on intermediate mass fragment (IMF) production, ever since the observation
of power laws in fragment mass-yield distributions \cite{Finn82zl,Minich82tk},
reminiscent of the condensation of liquid drops in a critical vapour
\cite{Fish67:droplet}. This effort was encouraged by the failure
of statistical models of hot nuclear decay \cite{Friedman83:stat-model,Cha88:Gemini,Dur92:simon}
to explain the observed fragment yields \cite{Bowman91:xe+au_50amev,Llope95:autocorrelations_imf_mult}
unless they suppose the break up of the system at low densities \cite{Friedman90:EES,MMMC:main-ref,SMM:main-ref}
where the phase transition may occur \cite{Ber83:spinodal}.

Signals of the phase transition in experimental data on multifragmentation
may be revealed by anomalously large fluctuations of fragment observables
\cite{Campi:perco,Mekjian:critical,Richert:2000hp}. The main obstacles
to such endeavours insofar as experimental data are concerned are
the huge statistical fluctuations inherent to small systems such as
atomic nuclei, and the still-open question of the mechanism(s) of
fragment production in heavy ion collisions at intermediate energies
\cite{Mor93:multifragmentation,Lott:binary,Toke:imf_prod_neck_decay,Dem96:isospin-midrap-55amev,Kunde96:imf_prod_112_124_sn_40amev,Larochelle:mech_frag_prod,I9-Mar97,I10-Luk97,I16-Neb99,I17-Pla99,I23-Lef00,Djerroud:heavy_residues,I28-Fra01,I29-Fra01,I31-Bor01,I37-Bel02,I36-Col03,I41-Cha02}.
Indeed in all experimental studies of the question, at the same time
as one is searching for evidence of the phase transition of a piece
of hot nuclear matter \cite{I29-Fra01}, one is (implicitly or not)
obliged to elucidate the manner in which such an excited system may
be formed in the course of certain reactions \cite{I28-Fra01}. The
solidity of any experimental evidence for a link between multifragmentation
and the nuclear matter phase diagram will be undermined by any remaining
ambiguity about the dominant mechanism of fragment production in the
selected sample of experimental events, and it would be preferable
to be able to address the question of {}``phases'' in nuclear multifragmentation
independently of reaction mechanism.

It is for this reason that the theory of the universal character of
order parameter fluctuations in finite systems proposed by Botet and
P\l oszajczak \cite{Botet:1997uf,Botet:1999sg,Botet:2000dk,Botet:2000rx}
provides an attractive opportunity to address the question of phase
transitions induced by heavy-ion collisions in the least ambiguous
way possible. According to their work, it is possible to obtain pertinent
information on the relationship between the formation of clusters
in a system and the phase transition(s) of said system without needing
to characterise the state of the system under study to the extent
of, for example, supposing it to be in thermodynamical equilibrium
at the time of cluster formation. One needs only to study the properties
of the clusters: specifically, all pertinent information can be obtained
from a sufficiently precise measurement of the event-by-event distributions
of cluster multiplicity and the size of the largest cluster produced.
The attractiveness of such an approach in the domain of nuclear fragmentation
reactions is obvious, for the reasons given above: although the mechanism
of fragment production remains an open question, for several years
now large solid-angle multidetector arrays such as INDRA have provided
high-quality data on the multiplicity and the size (charge) of the
fragments produced in such reactions.

A first application of the universal fluctuations theory to INDRA
data for central collisions of Xe+Sn from 25 to 50\amev~was published
in \cite{Botet:2001qp}. In this paper we will present the results
of the same analysis applied to a wide range of systems measured with
INDRA. As in \cite{Botet:2001qp} we observe a signature compatible
with the existence of different regimes at different bombarding energies.
This behaviour disappears for less central collisions. By applying
the analysis to data for colliding systems with total mass number
between 73 and 394 we will show the mass dependence of the energy
of transition between the two regimes, as well as presenting a systematic
study of the universal scaling functions observed in the low- and
high-energy regimes.

\section{Universal fluctuations of the order parameter in finite systems\label{sec:Universal-fluctuations}}

Universal scaling laws of fluctuations (the $\Delta$-scaling laws)
have been derived for equilibrium systems, and shown to apply also
in certain out-of-equilibrium situations. In a system in which the
second-order critical behaviour can be identified, the relation between
order parameter, criticality and scaling law of fluctuations has been
established and the relation between the scaling function and the
critical exponents has been found. Details can be found in \cite{Botet:2000rx}.

Experimental observables that may be related to a critical order parameter
can be identified through their $\Delta$-scaling behaviour. The $\Delta$-scaling
is observed when two or more probability distributions $P_{N}[m]$
of the observable $m$ for a system of 'size' $N$ collapse onto a
single scaling curve $\Phi(z_{(\Delta)})$ independent of system size
when plotted in terms of the scaling variables:

\begin{equation}
<m>^{\Delta}P_{N}[m]=\Phi(z_{(\Delta)})=\Phi\left(\frac{m-<m>}{<m>^{\Delta}}\right)\label{eq:delta-scaling}\end{equation}
where $<m>$ is the mean value of the distribution $P_{N}[m]$ and
$\frac{1}{2}\leq\Delta\leq1$. $<m>$ plays the role of a scale parameter
and can replace $N$ as a measure of the size of the system. A less
strong (necessary but not sufficient) condition for $\Delta$-scaling
is that the variance of the distribution should scale with its mean
value as

\begin{equation}
\sigma^{2}\sim<m>^{2\Delta}\label{eq:kappa-plot}\end{equation}
so that in a log-log plot of $\sigma^{2}$ versus $<m>^{2}$ data
should fall on a straight line of slope $\Delta$.

The scaling law Equation \ref{eq:delta-scaling} with $\Delta=\frac{1}{2}$
is associated with low temperature ({}``ordered'') systems, or with
observables which are not related to an order parameter. Scaling with
$\Delta=1$ is seen at high temperature ({}``disordered'' system)
and also for critical systems. For $m$ to be an order parameter it
must exhibit a corresponding change of $\Delta$-scaling regime when
some suitable control parameter (e.g. available energy, temperature,
bond-breaking probability, etc.) is varied.

Here it is worth saying a word about the comparison of experimental
distributions using Eq.\ref{eq:delta-scaling} and the determination
of $\Delta$-scaling regimes for data. As the transformation from
experimental observable to the scaling variable $z_{(\Delta)}$ is
a linear transformation, the form of the $\Phi(z_{(\Delta)})$ distribution
is the same as that of $P_{N}[m]$. However, the presence of the exponent
$\Delta$ in the scale factor $<m>^{\Delta}$ means that two identical
distributions will appear different (their widths will be different)
if they are scaled using a value of $\Delta$ which is not the one
relating the mean values and variances of the two distributions \emph{via}
Eq.\ref{eq:kappa-plot}. For example, consider two Gaussian distributions
whose widths and mean values are related \emph{via} Eq.\ref{eq:kappa-plot}
with $\Delta=1$. In this case the use of Eq.\ref{eq:delta-scaling}
with, for example, $\Delta=\frac{1}{2}$ would not lead to a universal
scaling function: the widths of the two scaled distributions would
be different, and, because of the normalisation of $\Phi(z_{(\Delta)})$,
the height of the distributions would differ also. It should also
be noted that the relationship between the mean and the variance of
a set of distributions, i.e. the value of $\Delta$ if Eq.\ref{eq:kappa-plot}
holds, is quite independent of the form of the distributions: a given
value of $\Delta$ does not imply a certain type of distribution,
and \emph{vice versa}.

More detailed information on the state of the system and the fragmentation
process may be found in the form of the scaling functions $\Phi(z_{(\Delta)})$,
Eq. \ref{eq:delta-scaling}. For systems far from a critical point,
the central limit theorem tells us that for an observable $m$ being
the sum of uncorrelated random variables, one should observe asymptotically
a Gaussian distribution for fluctuations of $m$ about its mean value.
If on the other hand $m$ is an extremal value such as the largest
among a set of uncorrelated random variables, then asymptotically
its distribution should be that of Gumbel's first asymptote \cite{Gumbel:extremal}.
For critical systems such general results do not exist, as, due to
the presence of correlations at all length-scales, the order parameter
distribution must depend on the precise details of the interaction
in this case. An asymptotic form of the large-$m$ scaling function
tail was derived in \cite{Botet:2000dk}, where close to a critical
point the order parameter distribution was expected to fall off like
$\exp-m^{\hat{\nu}}$ with $3\leq\hat{\nu}\leq6$, meaning that large
deviations of the order parameter from its mean value are strongly
suppressed.

There are two generic families of fragment production scenarios for
which the second-order phase transition has been identified, with
two different order parameters. These are 

\begin{itemize}
\item the fragment multiplicity for \emph{fragmentation scenarios} such
as the fragmentation-inactivation binary (FIB) model \cite{Botet:1992qr}.
\item the size of the largest cluster or fragment in the so-called \emph{aggregation
scenarios} such as percolation or Fisher droplet models \cite{Botet:2000dk}.
\end{itemize}
Therefore the elimination of one of these two easily measurable experimental
quantities as not having the order parameter $\Delta$-scaling behaviour
described above should give important information on the fragment
production process, by allowing to exclude one of the two scenarios.

\section{Presentation of data}

\subsection{Experimental details\label{sub:Experimental-details}}

In order to study as exhaustively as possible the question of the
existence of an order parameter or other {}``phase-like'' behaviour
in heavy-ion collisions, we have profited from the wide range of very
high quality data which has been obtained with the INDRA 4$\pi$ array
\cite{I3-Pou95,I4-Ste95,I5-Pou95} at the GANIL (Caen) and GSI (Darmstadt)
accelerator facilities. This charged product detector covers about
90\% of the 4$\pi$ solid angle. The total number of detection cells
is 336 arranged according to 17 rings centred on the beam axis. The
first ring ($2^{o}-3^{o}$) was composed of 12 fast NE102/NE115 phoswich
detectors during the experiments at the GANIL facility. For the GSI
experiments these were replaced by 12 telescopes composed of a 300
$\mu$m thick silicon detector and a 14 cm thick CsI(Tl) scintillator.
Rings 2 to 9 cover the angular range from $3^{o}$ to $45^{o}$ and
are made of three detector layers: a low pressure gas-ionization chamber
(5cm of C$_{3}$F$_{8}$ at 20 to 50 mbar, depending on the experiment
and the polar angle), a 300 $\mu$m thick silicon detector and a 14
to 10 cm thick CsI(Tl) scintillator. The remaining 8 rings cover the
angular range from $45^{o}$ to $176^{o}$ and have two detection
layers: ionization chamber and 7.6 to 5 cm thick CsI(Tl) scintillator.
Fragments with $Z$ up to the charge of the projectile are identified
with unit resolution in the forward region, when they are stopped
in the scintillator detectors. Beyond $45^{o}$, the charge resolution
is one unit up to $Z=16$, and a few charges for larger $Z$. Over
the whole angular range, a very good isotope identification is obtained
for $Z=1$ to $Z=3$, except for particles with low energies where
ambiguities are unresolved.

The energy calibration of the CsI(Tl) scintillators was obtained for
light charged particles (LCP) by means of the elastic and inelastic
scattering of secondary LCP beams (p,d,t,$^{3}$He,$^{4}$He) produced
by the fragmentation of a 95\amev~$^{16}$O beam in a thick C target.
These particles were then momentum selected by the {}``alpha magnetic
spectrometer'' of GANIL and scattered in a C or Ta target installed
in the INDRA reaction chamber. A typical energy resolution was about
4\%. Typical identification thresholds are a few 100 keV for light
particles, 0.7\amev~for $Z=3$, and 1.4\amev~for $Z=35$. A complete
technical description of INDRA, its calibration and its electronics
can be found in \cite{I4-Ste95,I5-Pou95,I3-Pou95,I14-Tab99,I33-Par02,I34-Par02,I44-Trz03}.

\subsection{Overview of data\label{sub:Overview-of-data}}

\begin{figure}[htbp]
\includegraphics[%
width = \columnwidth]{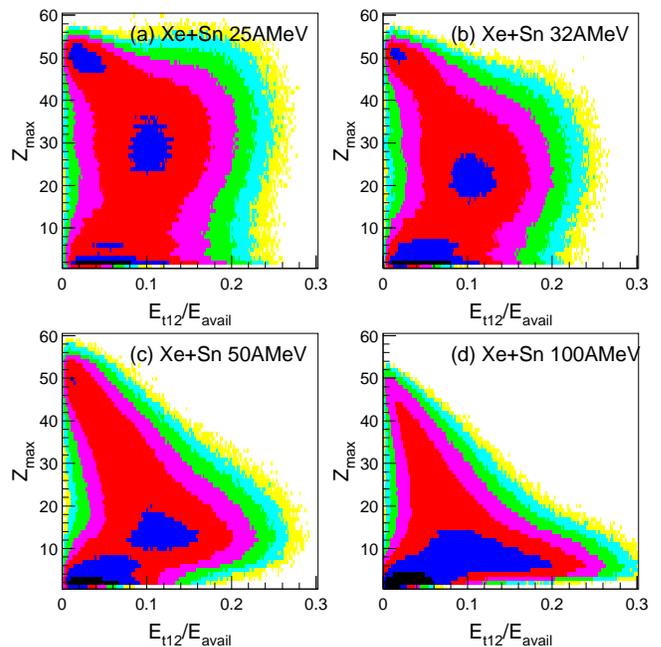}

\caption{(Color online) Experimental correlations between the size of the
largest detected fragment $Z_{max}$ and the total transverse energy
of light charged particles divided by the available c.m. energy, for
Xe+Sn collisions. A minimum of 4 well-identified charged particles
was required in the off-line analysis. Logarithmic contour levels
are shown corresponding to the number of events (darker tones indicate
larger numbers).\label{fig:zmax_vs_et12_xe+sn}}
\end{figure}

The data presented here cover a wide range of quasi-symmetric systems
studied with the INDRA array, with different total masses and bombarding
energies. They are: 

\begin{lyxlist}{00.00.0000}
\item [Ar+KCl]32, 40, 52 and 74\amev;
\item [Ni+Ni]32, 40, 52, 63, 74, 82 and 90\amev;
\item [Xe+Sn]25, 32, 39, 45, 50, 65{*}, 80{*} and 100{*}\amev;
\item [Au+Au]40{*}, 60{*} and 80{*}\amev.
\end{lyxlist}
The systems marked with an asterisk were measured during the experimental
campaign at the GSI facility. We have concentrated on symmetric colliding
systems in order to benefit from the maximum overall efficiency of
the INDRA array in this case. Before presenting the analysis of this
data set in terms of universal fluctuations we will give an overview
of the chief characteristics of these reactions.

\begin{figure}[htbp]
\includegraphics[%
width = \columnwidth]{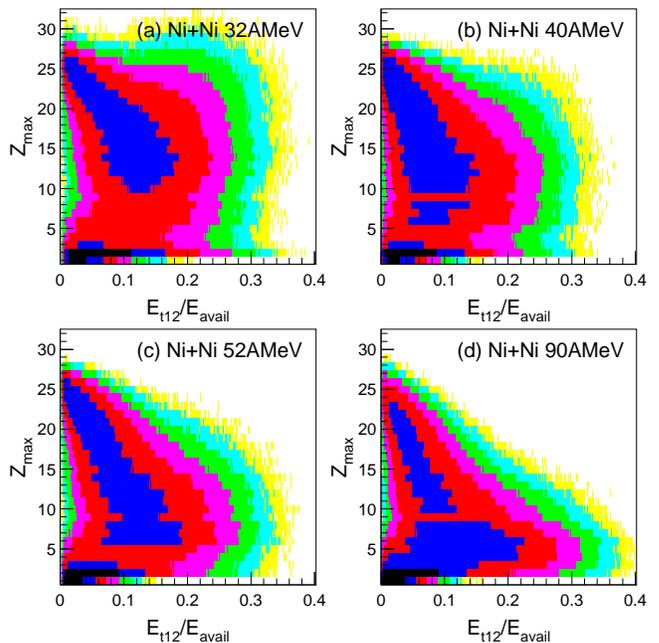}

\caption{(Color online) Experimental correlations between the size of the
largest detected fragment $Z_{max}$ and the total transverse energy
of light charged particles divided by available c.m. energy, for Ni+Ni
collisions. A minimum of 4 well-identified charged particles was required
in the off-line analysis. Logarithmic contour levels are shown corresponding
to the number of events (darker tones indicate larger numbers).\label{fig:zmax_vs_et12_ni+ni}}
\end{figure}

Figure \ref{fig:zmax_vs_et12_xe+sn} shows experimental data for the
Xe+Sn system. The contour plots show the number of events measured
corresponding to each value of the size of the largest detected fragment,
$Z_{max}$, and of the fraction of the available energy converted
into transverse energy of light charged particles, $E_{t12}/E_{avail}$.
This latter quantity has been shown \cite{I10-Luk97,I17-Pla99,I19-Met00,I21-Boc00}
to be principally related to the geometry of heavy-ion collisions
in this energy domain, and is particularly well-suited to sorting
events measured with the INDRA detector with little bias, because
the efficiency of the array for light charged particle detection is
$\approx90\%$ whatever the centrality/reaction mechanism. The data
shown were recorded with an on-line trigger requiring that at least
four detectors fired in coincidence, while in the off-line analysis
we required at least four correctly identified charged products in
each considered event. Very similar plots were obtained for all other
data studied in this paper, as for example the data for the Ni+Ni
system shown in Figure \ref{fig:zmax_vs_et12_ni+ni}. They give an
overview of the evolution of reaction mechanisms with beam energy
and impact parameter.

For the least violent collisions (small $E_{t12}/E_{avail}$) two
strong contributions to the total cross-section are observed, depending
on whether the projectile-like fragment was detected ($Z_{max}\approx Z_{proj}$)
or not ($Z_{max}<10$). For most of these collisions the target-like
residue is too slow-moving to be detected and/or correctly identified
by the INDRA array. With increasing beam energy, the contribution
from projectile-like fragments appears to decrease in importance.
This is due to the increasingly forward-focused kinematics of the
reactions and the smaller grazing angle at higher energy, which mean
that projectile-like fragments from peripheral collisions are more
likely to be deflected to angles too small for them to be detected
in the first ring of the INDRA array (see \ref{sub:Experimental-details}). 

For more central collisions (larger $E_{t12}/E_{avail}$) the outcome
of the reaction depends on bombarding energy and the mass of the colliding
nuclei. For the Xe+Sn system at the lowest incident energies (Figures
\ref{fig:zmax_vs_et12_xe+sn}a-b), the cross-section for all but the
most peripheral collisions is dominated by events for which the heaviest
detected fragment has around one half of the charge of the projectile
nucleus, and there is a discontinuous evolution in cross section between
the most peripheral collisions and these events. This is due to the
onset of fission of projectile-like nuclei above a certain threshold
excitation energy which is attained in mid-peripheral and more central
reactions \cite{I21-Boc00}. At higher bombarding energies (Figures
\ref{fig:zmax_vs_et12_xe+sn}c-d) a more continuous evolution of projectile-like
residue size with the collision violence is seen. For a lighter system
such as Ni+Ni (Figures \ref{fig:zmax_vs_et12_ni+ni}a-b) a continuous
ridge goes from the most peripheral toward more central collisions
showing that the size of the projectile-like residues decrease continuously
with decreasing impact parameter, as fission is not a predominant
decay channel even at high excitation energies for such light nuclei.
The correlations between $Z_{max}$ and $E_{t12}/E_{avail}$ for the
Ar+KCl system (not shown) are very similar to those for Ni+Ni, whereas
for the Au+Au system they resemble those for Xe+Sn at 50 and 100\amev~except
that for the most peripheral collisions a clear contribution from
the fission of gold quasiprojectiles is seen. 

Finally let us remark the similarity between Figures \ref{fig:zmax_vs_et12_xe+sn}a
and \ref{fig:zmax_vs_et12_ni+ni}a, concerning the production of heavy
residues at low incident energy. Both figures show that there are
two distinct contributions to the total yield for residues with $Z$
close to that of the projectile. The first was discussed in the previous
paragraph and is due to projectile-like fragments produced in peripheral
collisions (small $E_{t12}/E_{avail}$ values). This is the dominant
contribution. Nevertheless there is a second contribution which is
associated with a broad distribution of large residue sizes $Z_{proj}/2<Z_{max}\lesssim Z_{proj}$
as well as an equally broad distribution of $E_{t12}/E_{avail}$ values
corresponding to mid- to central collisions. The appearance of these
experimental correlations suggests that the mean value of the charge
of the heaviest detected fragment, $<Z_{max}>$, first decreases with
increasing collision violence and then increases for the most {}``central''
\begin{figure}[htbp]
\includegraphics[%
width = \columnwidth]{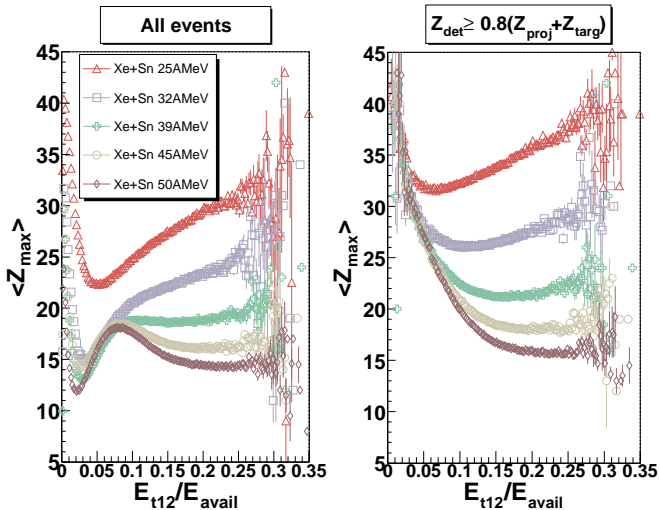}

\caption{(Color online) The mean charge of the largest fragment detected in
each event, $Z_{max}$, as a function of the total transverse energy
of light charged particles (normalised to the available centre of
mass energy) for Xe+Sn collisions between 25 and 50\amev. Vertical
bars show the estimated statistical error on the mean. It should be
noted that events with $E_{t12}/E_{avail}\gtrsim\textrm{0.2}$ correspond
to only 1\% of the total measured cross-section.\label{fig:mean_zmax_vs_et12}}
\end{figure}
collisions at the lowest beam energy for these two systems. 

According to Figures \ref{fig:zmax_vs_et12_xe+sn} and \ref{fig:zmax_vs_et12_ni+ni}
such reactions make a relatively important contribution to the production
of heavy residues in central collisions of Ni+Ni and Xe+Sn at 32 and
25\amev~respectively. When the incident energy increases (40\amev~for
Ni+Ni, Figure \ref{fig:zmax_vs_et12_ni+ni}b, and 32\amev~for Xe+Sn,
Figure \ref{fig:zmax_vs_et12_xe+sn}b) the size and yield of the associated
residues decrease making this contribution harder to distinguish.
From Figures \ref{fig:zmax_vs_et12_ni+ni}c-d and Figures \ref{fig:zmax_vs_et12_xe+sn}c-d
it appears that the cross-sections associated with these reactions
become very small above $\sim40$\amev, for the Ni+Ni system, or
above $\sim32$\amev, for the Xe+Sn system (for the Ar+KCl system,
not shown here, this contribution is discernible up to 52\amev).
However, more sensitive analyses (e.g. see below for Xe+Sn) can reveal
the survival of such reactions at higher incident energies.

Figure \ref{fig:mean_zmax_vs_et12} shows the evolution of $<Z_{max}>$
with collision centrality for the Xe+Sn system between 25 and 50\amev.
It can be seen that the mean charge of the heaviest fragment produced
in each event increases with increasing centrality for 25, 32, and,
very slightly, 39\amev. It should be noted that this tendency is
observed whether one considers all (Figure \ref{fig:mean_zmax_vs_et12}
left panel) or only well-measured (right panel - ratio of total detected
charge, $Z_{det}$, to total charge of projectile and target required
to be at least 80\%) events. The requirement of well-measured events
reveals the monotonous decrease of the mean charge of projectile-like
fragments in peripheral collisions ($E_{t12}/E_{avail}<0.1$) by excluding
from the average those events where neither target-like nor projectile-like
fragments were detected (compare Figure \ref{fig:zmax_vs_et12_xe+sn}).

\begin{figure}[htbp]
\includegraphics[%
width = \columnwidth]{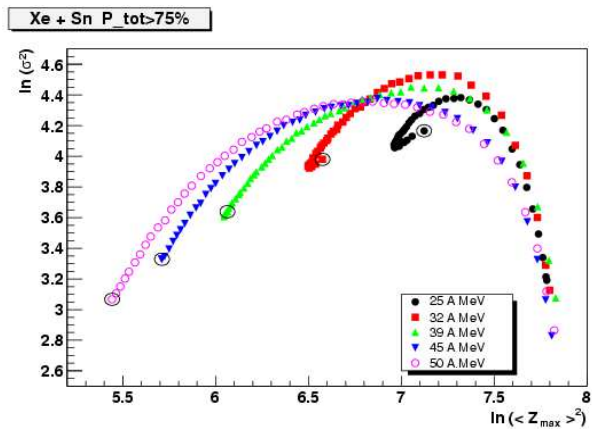}

\caption{(Color online) Collisions from the most peripheral with detection
of a quasi-projectile residue (furthest right-hand side of the figure)
to the most central (indicated by a ring) are classed into bins corresponding
to 1\% of the total measured cross section.\label{fig:mean_and_variance_zmax_all_collisions}}
\end{figure}

Another way of excluding poorly-measured events is to keep only those
for which most of the momentum of the incident beam particles is reconstructed
from the detected nuclei. This gate on data has the effect of retaining
all events for which at least a good reconstruction of the quasi-projectile
residue and products was obtained, for the most peripheral collisions,
whilst in the most central collisions we achieve an almost complete
reconstruction of the event. In this way we can follow, in Figure
\ref{fig:mean_and_variance_zmax_all_collisions}, the correlated evolution
of the variance and the mean of the $Z_{max}$ distribution as a function
of collision violence: from peripheral collisions leading to a slightly
excited quasi-projectile (large $<Z_{max}>$ and small variance) to
the most central collisions (indicated by a ring around the last data
point for each beam energy). Each point in this figure corresponds
to $E_{t12}$ cuts defined by slicing the minimum-bias distribution
into 100 bins, each containing an equal number of events. Thus each
point represents 1\% of the measured cross-section, although due to
our requirement of well-measured events the most peripheral (least
well measured) events are absent. It allows to see how in the most
central collisions two different behaviours are observed depending
on the incident beam energy: below 39\amev~in the 3-5\% most central
collisions the mean charge of the largest fragment actually begins
to increase again, whilst the correlation between this mean value
and the variance of the distribution is not the same as for the majority
of other impact parameters. 

This figure suggests that at beam energies $\lesssim$39\amev~the
origin of the heaviest fragment is not the same in central collisions
as for the rest of the reactions. The increase of $<Z_{max}>$ for
the most central collisions at these energies suggests a contribution
from (incomplete) fusion reactions where the heaviest fragment detected
is an evaporation residue. The disappearance of this phenomenon for
$\gtrsim$39\amev~can be interpreted as signalling the disappearance
of fusion-evaporation residues, and marking the onset either of fusion-multifragmentation
\cite{I28-Fra01} or of transparency in central collisions (incomplete
stopping) \cite{I16-Neb99,Ono02:wpshrink,LeFevre:2003sx}.

\section{Analysis of data in terms of universal fluctuations\label{sec:Analysis-of-data}}

\subsection{Selection of central collisions}

We begin our analysis by extending the results of \cite{Botet:2001qp},
for central collisions of Xe+Sn, to a wider range of systems measured
with the INDRA multidetector array. As was discussed in Sec. \ref{sec:introduction},
the analysis of universal fluctuations in multifragmentation data
does not require to know (or to suppose) how fragments are formed.
However the comparison of events with very different collision geometries
is unlikely to give meaningful results. Therefore we will limit our
study to very central collisions for which the geometrical overlap
between projectile and target is as close as possible to total. The
need to select equivalent classes of events for a large range of system
masses and beam energies with as little dependence as possible on
detector efficiency led us to use the total transverse energy of light
charged particles, $E_{t12}$. 

The sorting variable $E_{t12}$ has been studied specifically for
the INDRA detector \cite{I17-Pla99,Frankland:2002aq}. It was shown
that, for a given projectile-target system, the minimum-bias inclusive
distributions of $E_{t12}$ for different beam energies scale as a
function of the available centre of mass energy, consistent with $E_{t12}$
being mainly sensitive to the geometry of the collisions, i.e. that
it is a good indicator of collision centrality. Moreover, as it was
pointed out in Sec. \ref{sub:Overview-of-data}, the efficiency of
the INDRA detector for light charged particles is almost independent
of the type of reaction under study, allowing to use this variable
to sort all data considered in this paper in the same way.

In addition, in this way we avoid any direct link between the studied
observables (size of the largest fragment in each event and fragment
multiplicity) and the variable used for the selection of events. We
are therefore in the best possible situation to avoid distortion of
the data due to autocorrelations with the experimental filter \cite{Llope95:autocorrelations_imf_mult}. 

The data analysed in the following with the largest values of $E_{t12}$
correspond to 1\% of the total number of events recorded during the
experiment with a minimum-bias condition ($b<0.1b_{max}$ in the geometrical
approximation of \cite{Cavata:ip} where the maximum impact parameter
$b_{max}$ is smaller than the sum of the radii of projectile and
target due to the experimental trigger condition). We were able to
check whether the largest detected charged fragment of each event
is really the largest, using the total detected charge of the event,
$Z_{tot}$.

\subsection{Establishing an {}``order parameter'' for nuclear multifragmentation
with no model-dependent hypotheses\label{sub:order-parameter}}

\begin{figure}[htbp]
\includegraphics[%
width = \columnwidth]{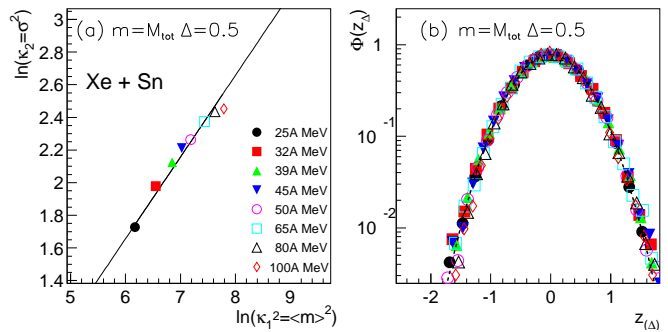}

\caption{(Color online) Data for the total multiplicity of charged products
with $Z\geq1$, $M_{tot}$, for well-measured central collisions of
Xe+Sn ($b<0.1b_{max}$, $Z_{det}\geq0.8(Z_{proj}+Z_{targ})$). (a)
Log-log plot of the variance versus the squared mean value of the
distribution of $M_{tot}$ for each bombarding energy. The straight
line represents Eq. \ref{eq:kappa-plot} with $\Delta=\frac{1}{2}$.
(b) Experimental distributions of $M_{tot}$ expressed in the variables
of the second scaling law, i.e. Equation \ref{eq:delta-scaling} with
$\Delta=\frac{1}{2}$. The dashed curve is a Gaussian fit to all the
data.\label{fig:xe+sn_25-100_mtot_delta=3D1/2}}
\end{figure}

Generic models of cluster production may be classed into two types.
The first, the class of \char`\"{}fragmentation\char`\"{} scenarios
in which a system is broken up by a series of binary splittings or
some other physical process, has for its order parameter the number
or multiplicity of clusters. For such models, it is the multiplicity
which exhibits different $\Delta$-scaling regimes if the system has
different phases. The second class of models, in which clusters are
built up by \char`\"{}aggregation\char`\"{} of smaller constituents,
has for its order parameter the size of the largest cluster. As INDRA
does not measure the mass but only the atomic number of fragments,
we will assume in the following that the largest fragment of each
event corresponds to $Z_{max}$, the fragment with the largest atomic
number.

Figure \ref{fig:xe+sn_25-100_mtot_delta=3D1/2} shows the data for
the total multiplicity of charged products with $Z\geq1$, $M_{tot}$,
for central collisions of Xe+Sn from 25 to 100\amev. In Figure \ref{fig:xe+sn_25-100_mtot_delta=3D1/2}a
we plot the natural logarithm of the variance of the measured $M_{tot}$
distributions as a function of the natural logarithm of the square
of the mean value. The estimated statistical errors of these quantities
are smaller than the symbols used. If fluctuations obey the universal
scaling law Eq. \ref{eq:kappa-plot}, then the data must fall on a
straight line of slope $\Delta$ in this plot. It can be seen in Figure
\ref{fig:xe+sn_25-100_mtot_delta=3D1/2}a that this is true to a fair
approximation, and that the variance of the multiplicity distribution
grows with increasing bombarding energy as $<M_{tot}>$.

\begin{figure}[htbp]
\includegraphics[%
width = \columnwidth]{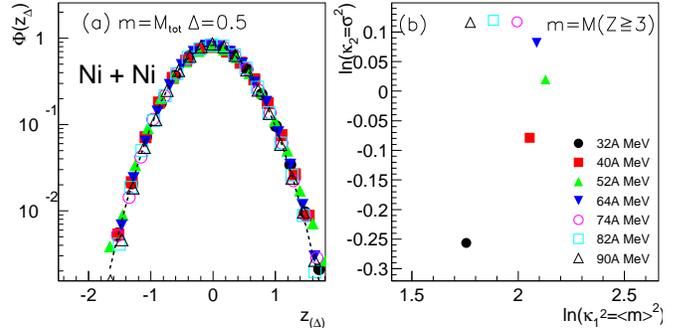}

\caption{(Color online) Data for well-measured central collisions of Ni+Ni
($b<0.1b_{max}$, $Z_{det}>0.8(Z_{proj}+Z_{targ})$). (a) Distributions
of $M_{tot}$ expressed in the variables of the second scaling law,
i.e. Equation \ref{eq:delta-scaling} with $\Delta=\frac{1}{2}$.
The dashed curve refers to a global Gaussian fit to the data (dashed
curve). (b) Log-log plot of the variance versus the squared mean value
of the distribution of fragment multiplicity, $M_{Z\geq3}$, for each
bombarding energy.\label{fig:ni+ni_32-90_mtot_mimf_delta=3D1/2}}
\end{figure}

Figure \ref{fig:xe+sn_25-100_mtot_delta=3D1/2}b shows that the multiplicity
distributions for different bombarding energies collapse to a unique
distribution (scaling function) when expressed in terms of the second
scaling law (Equation \ref{eq:delta-scaling} with $\Delta=\frac{1}{2}$),
as suggested by Figure \ref{fig:xe+sn_25-100_mtot_delta=3D1/2}a.
It should be noted that examination of an observable's scaling properties
in this way is far more constraining than that of Fig. \ref{fig:xe+sn_25-100_mtot_delta=3D1/2}a,
when the statistics of the data samples allow it. The observed scaling
function is very well approximated by a Gaussian distribution (dashed
curve on the figure). The scaling properties of total multiplicity
fluctuations are therefore the same for all bombarding energies in
the range 25--100\amev. This is true not only for the Xe+Sn data
but also for all the data we have studied in this paper: see for example
the data for $^{58}$Ni+$^{58}$Ni collisions in Figure \ref{fig:ni+ni_32-90_mtot_mimf_delta=3D1/2}a.
Therefore the total multiplicity of charged particles for central
collisions in this energy range does not show any evidence of 'anomalous'
or 'critical' behaviour.

\begin{figure}[htbp]
\includegraphics[%
width = \columnwidth]{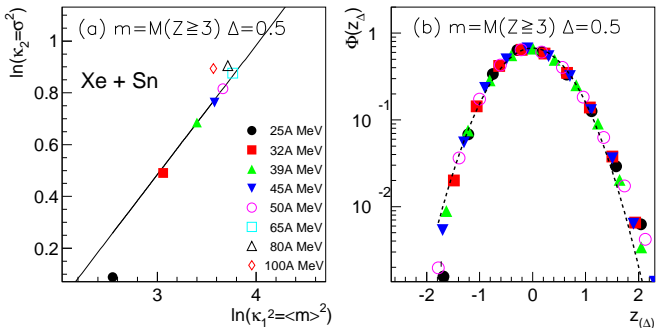}

\caption{(Color online) As Figure \ref{fig:xe+sn_25-100_mtot_delta=3D1/2}
but now for the multiplicity of fragments with $Z\geq3$, $M_{Z\geq3}$.
(b) Data for bombarding energies 32--65\amev~expressed in the variables
of the second scaling law, i.e. Equation \ref{eq:delta-scaling} with
$\Delta=\frac{1}{2}$.\label{fig:xe+sn_25-100_mimf_delta=3D1/2}}
\end{figure}

The multiplicity $M_{tot}$ is dominated by the multiplicity of light
charged particles, $M_{LCP}$, which is typically 3--4 times greater
than the multiplicity of IMF ($Z\geq3$). However, similar conclusions
as for $M_{tot}$ can be drawn for the IMF multiplicity, $M_{Z\geq3}$,
for which the Xe+Sn data are presented in Figure \ref{fig:xe+sn_25-100_mimf_delta=3D1/2},
the Ni+Ni data in Figure \ref{fig:ni+ni_32-90_mtot_mimf_delta=3D1/2}b.
Figure \ref{fig:xe+sn_25-100_mimf_delta=3D1/2}a shows that the widths
of $M_{Z\geq3}$ distributions for Xe+Sn at bombarding energies 32
to 65\amev~increase with the mean multiplicity according to a $\Delta=\frac{1}{2}$
scaling law, and this is confirmed by Figure \ref{fig:xe+sn_25-100_mimf_delta=3D1/2}b.
The width for the 25\amev~system falls below this {}``systematic''
trend, which may indicate that this energy is close to the threshold
for multifragmentation in central collisions (the mean multiplicity
for this system is approximately 3 fragments with $Z\geq3$), leading
to reduced fluctuations of the fragment multiplicity. Recent data
obtained with INDRA for the same system at bombarding energies from
8 to 20\amev~will allow to study this point in more detail.

Nevertheless the multiplicity of fragments does show some slightly
more interesting features than $M_{tot}$: let us remark the {}``back-bending''
in Figure \ref{fig:xe+sn_25-100_mimf_delta=3D1/2}a for bombarding
energies $>65$\amev, indicating the decrease of mean fragment multiplicity
at the highest bombarding energies. The same behaviour is seen more
clearly in Figure \ref{fig:ni+ni_32-90_mtot_mimf_delta=3D1/2}b for
Ni+Ni collisions above 52\amev. Let us note in passing that the available
energy for the maximum of fragment production in the Ni+Ni system
is here much lower (13\amev) than that reported in \cite{Sisan01:imf_emission}
(17.5\amev).

This type of {}``rise and fall'' behaviour has been observed in
many different data sets \cite{Ogi91:rise_and_fall,Tsang93:onset_of_vaporization,Peaslee94:rise_and_fall_kr+au,Sisan01:imf_emission}.
The absolute value of the maximum mean multiplicity and the energy
at which it occurs in central collisions are not only system-dependent
\cite{Sisan01:imf_emission} but also selection-dependent \cite{Llope95:autocorrelations_imf_mult}
and detector-dependent. It should be noted that in the framework of
the universal fluctuations theory this rise and fall is not consistent
with the fragment multiplicity being an order parameter. Rather, in
the energy domain for which fluctuations of $M_{Z\geq3}$ show a regular
scaling behaviour (32--65\amev~for Xe+Sn), it can be seen that this
scaling is always of the second kind ($\Delta=\frac{1}{2}$) and in
this domain the multiplicity distributions all collapse to a unique,
quasi-Gaussian scaling function (Figure \ref{fig:xe+sn_25-100_mimf_delta=3D1/2}b). 

Therefore, neither the total charged multiplicity nor the fragment
multiplicity have fluctuations which indicate the presence of different
{}``phases'' for central collisions in this energy range. It should
be noted, however, that we cannot exclude the possibility of a different
behaviour of observables to which we do not have access in this data,
for example the true total multiplicity including neutrons. Neither
does our result exclude the possibility that in other fragmentation
reactions, e.g. at higher bombarding energies and/or in spectator
decays rather than central collisions, the total or fragment multiplicity
may be the pertinent order parameter. For our data on central collisions
between 25 and 100\amev~however we will from now on only consider
the other possible order parameter for fragmentation, the size of
the largest fragment.

\begin{figure}[htbp]
\includegraphics[%
width = \columnwidth]{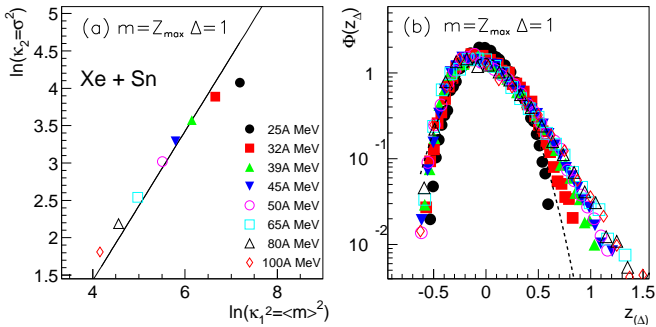}

\caption{(Color online) As Figure \ref{fig:xe+sn_25-100_mtot_delta=3D1/2}
(well-measured central Xe+Sn collisions) but for the charge $Z_{max}$
of the heaviest fragment detected event-by-event, and $\Delta=1$.\label{fig:xe+sn_25-100_zmax_delta=3D1}}
\end{figure}

Looking at the log-log plot of the first two cumulant moments of $Z_{max}$
(Figure \ref{fig:xe+sn_25-100_zmax_delta=3D1}a) we can suspect some
evolution of the scaling behaviour of this observable's fluctuations
with increasing beam energy: the data do not appear to fall on a single
straight line, but rather seem to be grouped into two {}``branches''
with different slopes. Although most of the data in Figure \ref{fig:xe+sn_25-100_zmax_delta=3D1}a
lie close to a line of slope $\Delta=1$, this does not correspond
to a universal scaling law, as all data in Figure \ref{fig:xe+sn_25-100_zmax_delta=3D1}b
do not collapse onto a single universal curve under the first scaling
law (Equation \ref{eq:delta-scaling} with $\Delta=1$). This confirms
that there is a change in the scaling behaviour of $Z_{max}$ fluctuations
with increasing energy. The charge of the largest fragment in each
event, and not the fragment multiplicity, seems therefore to be a
good candidate for the order parameter of multifragmentation in central
collisions at these energies.

\subsection{Evolution of the scaling behaviour of order parameter fluctuations
with incident energy\label{sub:scaling-with-energy}}

\begin{figure}[htbp]
\includegraphics[%
width = \columnwidth]{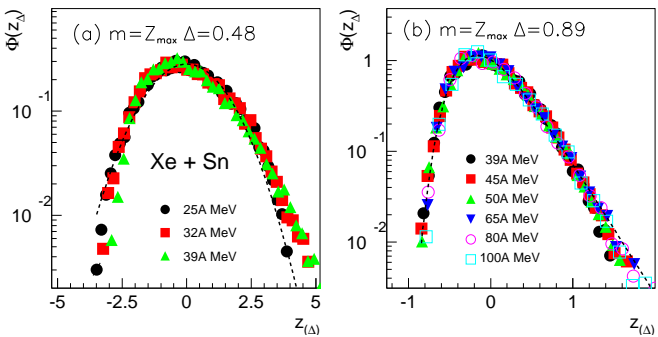}

\caption{(Color online) (a) $Z_{max}$ distributions for Xe+Sn collisions
at 25--39\amev, scaled according to Equation \ref{eq:delta-scaling}
with $\Delta$ given by a linear fit to the data of Figure \ref{fig:xe+sn_25-100_zmax_delta=3D1}a
of the form $\sigma\sim<m>^{\Delta}$. The dashed curve is a best
fit to scaled data using a Gaussian distribution. (b) As (a) but for
bombarding energies 39--100\amev. The dashed curve is a best fit
to scaled data using the Gumbel distribution, Eq.\ref{eq:gumbel}.\label{fig:xe+sn_zmax_delta=3D1/2_delta=3D1}}
\end{figure}

Figure \ref{fig:xe+sn_25-100_zmax_delta=3D1}b shows that the scaled
$Z_{max}$ distributions for Xe+Sn collisions from 45\amev~upwards
are nearly identical even in the large-$Z_{max}$ tails which are
2 orders of magnitude less probable than the most probable value of
$Z_{max}$. The 39\amev~data can also be included in this group
if the small differences in the tail of the distribution are neglected.
However, the 32\amev~distribution is clearly significantly narrower,
while the distribution for 25\amev~is evidently of a different form,
as can be seen both in the tails and around the maximum (see comments
on the comparison of experimental distributions with this technique
in Sec.\ref{sec:Universal-fluctuations}).

Using the values of $\Delta$ extracted from Fig.\ref{fig:xe+sn_25-100_zmax_delta=3D1}a
using a linear fit to the two 'branches' (25-39\amev~and 39-100\amev)
we find a good scaling of distributions using $\sigma\sim<Z_{max}>^{1/2}$
at the lowest beam energies (Figure \ref{fig:xe+sn_zmax_delta=3D1/2_delta=3D1}a),
whereas for the higher energies the fluctuations of the size of the
largest fragment increase like $\sigma\sim<Z_{max}>$ (Figure \ref{fig:xe+sn_zmax_delta=3D1/2_delta=3D1}b).
An approximately equally good scaling for 39\amev~data is achieved
in both cases, and this energy may be considered as close to a transition
between the two regimes. Figure \ref{fig:xe+sn_zmax_delta=3D1/2_delta=3D1}
also shows that the shape of the scaling function changes with increasing
beam energy: it is nearly Gaussian at low energy (the dashed curved
in Figure \ref{fig:xe+sn_zmax_delta=3D1/2_delta=3D1}a represents
a best fit to all data with a Gaussian distribution), but at higher
energies it is rather asymmetric with a near-exponential tail for
$Z_{max}$ greater than its most probable value (see Sec. \ref{sub:form}).

These observations establish the size (or charge) of the largest fragment
as the most likely {}``order parameter'' for fragment production
in central collisions in this energy range. They also show that data
can be assigned to one of two regimes depending on the $\Delta$-scaling
properties of this order parameter. At low energies systems obey the
second-scaling law associated with an ordered phase, while at higher
energies the first-scaling law is observed, typical of a disordered
phase (large fluctuations) \cite{Botet:2000dk}. It should be noted
that these regimes are defined, not by the average size of the largest
fragment produced for a given incident energy (which would be a typical
way to define e.g. liquid and gas phases), but rather by the way in
which the fluctuations of this quantity evolve compared to its mean
value. The identification of the largest fragment size as order parameter
also indicates some similarity between fragment production in central
collisions and the aggregation models with the second-order phase
transition mentioned in Sec. \ref{sec:Universal-fluctuations}, whereas
the monotonous behaviour of total and fragment multiplicities excludes
{}``shattering'' fragmentation models such as FIB \cite{Botet:1992qr}. 

\begin{figure}[htbp]
\includegraphics[%
width = \columnwidth]{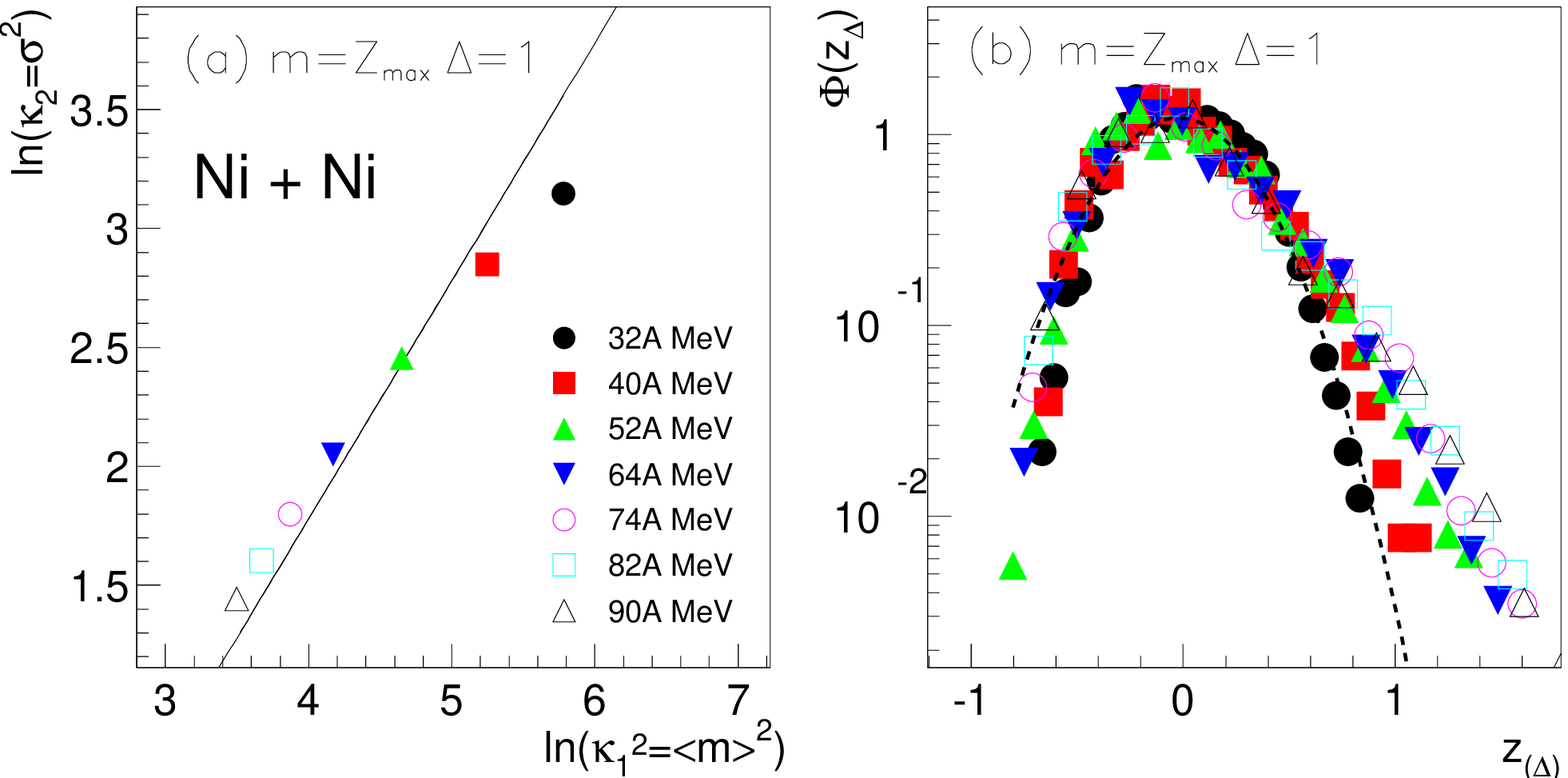}

\caption{(Color online) As Figure \ref{fig:xe+sn_25-100_zmax_delta=3D1} but
for collisions of Ni+Ni from 32 to 90\amev~bombarding energy.\label{fig:ni+ni_32-90_zmax_delta=3D1}}
\end{figure}

Another, not necessarily incompatible, interpretation of the different
$\Delta$ regimes is suggested by Fig. \ref{fig:mean_and_variance_zmax_all_collisions}.
As we remarked in Sec. \ref{sub:Overview-of-data} this figure suggests
that at beam energies $\lesssim$39\amev~the heaviest fragment in
central collisions is an evaporation residue of an (incomplete) fusion
reaction. The data presented in Fig. \ref{fig:xe+sn_25-100_zmax_delta=3D1}a
for incident energies 25-50\amev~correspond to the points highlighted
by a circle in Fig. \ref{fig:mean_and_variance_zmax_all_collisions}.
Therefore the transition from a $\Delta\sim\frac{1}{2}$ to a $\Delta\sim1$
branch can be interpreted as being linked to the disappearance of
fusion-evaporation residues, and signalling either the onset of fusion-multifragmentation
(phase transition) or the onset of transparency in central collisions
(incomplete stopping).

It should be noted that the transition energy of around 39\amev~for
the Xe+Sn system is slightly higher than that found in \cite{Botet:2001qp}
(32\amev). This is due to the different method of selection of 'central
collisions'. In this paper we define central collisions based on the
amount of energy deposited into the light charged particles' degrees
of freedom ($E_{t12}$). As we are dealing with finite systems this
reduces the energy available for the fragment degrees of freedom,
due to energy conservation. This is consistent with the fact that
when central collisions are selected based on fragment degrees of
freedom (the fragment kinetic energy flow angle selection of \cite{Lec94:theta_flow,I9-Mar97,I28-Fra01})
the multifragmentation regime is observed at lower incident energy
for the same system (32\amev~for Xe+Sn in \cite{I29-Fra01}), as
is the transition to the $\Delta=1$ scaling regime. 

\begin{figure}[htbp]
\includegraphics[%
width = \columnwidth]{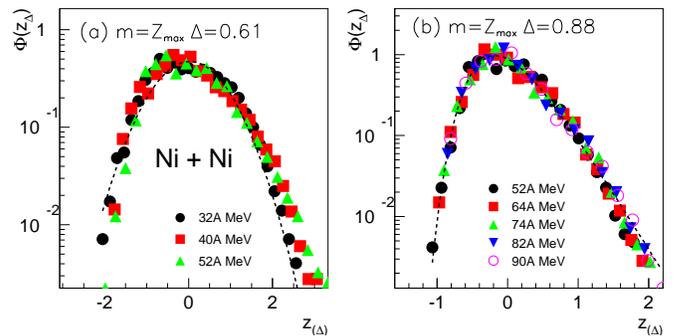}

\caption{(Color online) (a) $Z_{max}$ distributions for Ni+Ni collisions
at 32--52\amev, scaled according to Equation \ref{eq:delta-scaling}
with $\Delta$ given by a linear fit of the form $\sigma\sim<m>^{\Delta}$.
The dashed line is a best fit using a Gaussian distribution. (b) As
(a) but for energies 52--90\amev. The dashed line is a best fit using
the Gumbel distribution, Eq.\ref{eq:gumbel}.\label{fig:ni+ni_zmax_delta=3D1/2_delta=3D1}}
\end{figure}

The observed scaling properties of $Z_{max}$ fluctuations are confirmed
by the Ni+Ni data, as shown by Figure \ref{fig:ni+ni_32-90_zmax_delta=3D1}
and Figure \ref{fig:ni+ni_zmax_delta=3D1/2_delta=3D1}. Most data
follow a $\Delta\approx1$ scaling law for $Z_{max}$ fluctuations,
except at the lowest energies. The data do not all collapse to a single
scaling function in terms of a $\Delta=1$ law (Figure \ref{fig:ni+ni_32-90_zmax_delta=3D1}b),
but only for bombarding energies $E\geq52$\amev~(Figure \ref{fig:ni+ni_zmax_delta=3D1/2_delta=3D1}b).
The similarity between the scaling functions observed for the two
different systems should be noted, both in the {}``ordered'' ($\Delta\sim\frac{1}{2}$)
and {}``disordered'' ($\Delta\sim1$) regimes (Figures \ref{fig:xe+sn_zmax_delta=3D1/2_delta=3D1}
\& \ref{fig:ni+ni_zmax_delta=3D1/2_delta=3D1} and Table \ref{tab:scaling_function_parameters}). 

\subsection{System-size dependence of energy of transition from {}``ordered''
to {}``disordered'' regime\label{sub:scaling-with-size}}

\begin{figure}[htbp]
\includegraphics[%
width = \columnwidth]{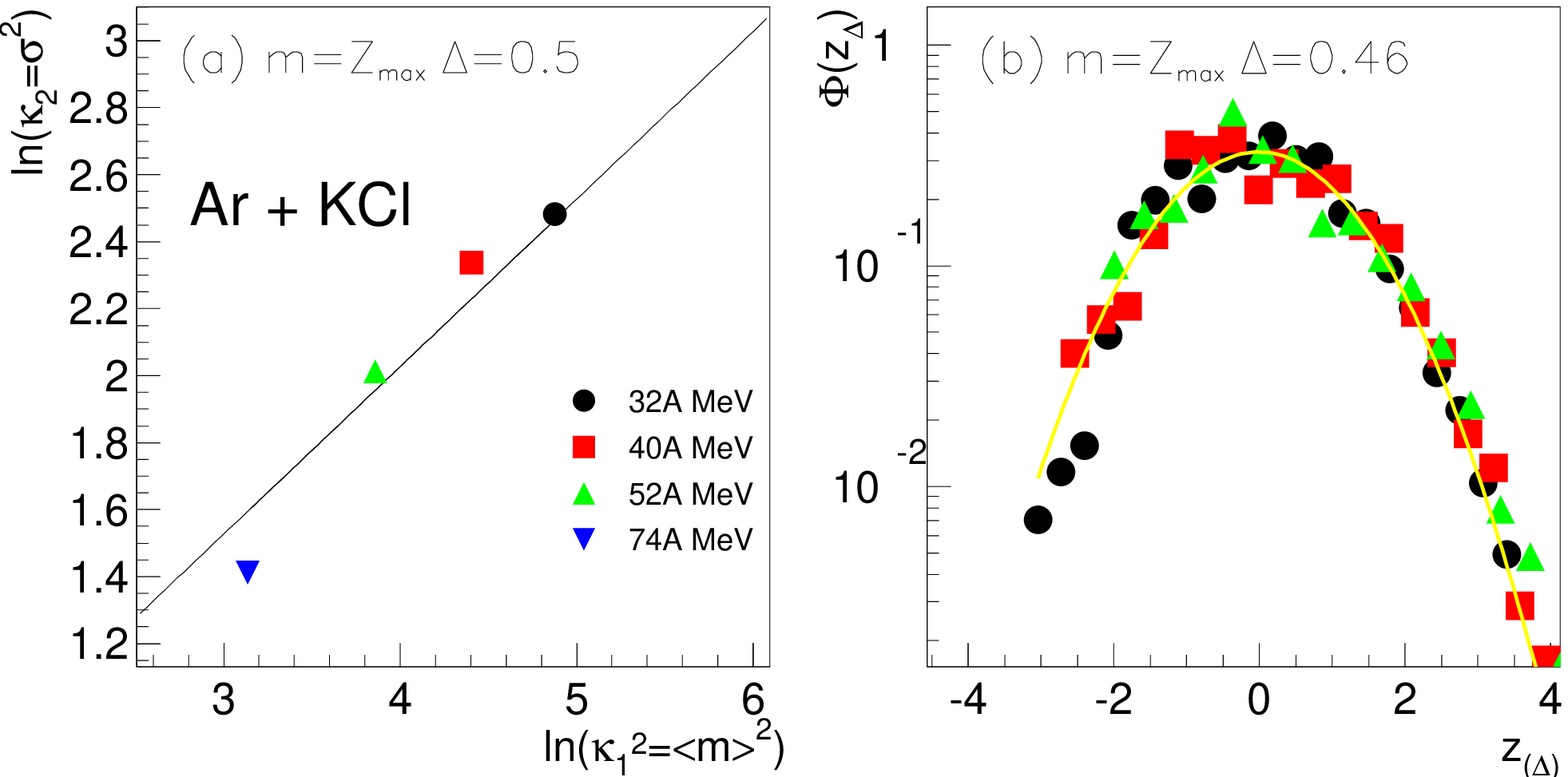}

\caption{(Color online) As Figure \ref{fig:xe+sn_25-100_zmax_delta=3D1} but
for collisions of Ar+KCl from 32 to 74\amev~bombarding energy.\label{fig:ar+kcl_32-74_zmax_delta=3D1/2}}
\end{figure}

\begin{figure}[htbp]
\includegraphics[%
width = \columnwidth]{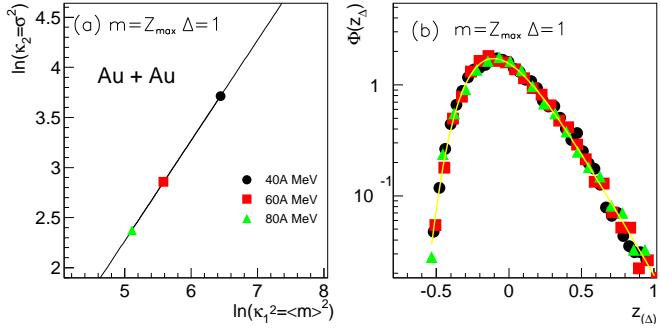}

\caption{(Color online) As Figure \ref{fig:xe+sn_25-100_zmax_delta=3D1} but
for collisions of Au+Au from 40 to 80\amev~bombarding energy.\label{fig:au+au_40-80_zmax_delta=3D1}}
\end{figure}

We observe for the Xe+Sn and Ni+Ni data that the energy ranges corresponding
to the different regimes are not the same for two systems of different
total mass, the {}``transition'' occurring around 39\amev~for
Xe+Sn and 52\amev~for Ni+Ni. The {}``disordered'' regime begins
at lower incident (or available) energy for the heavier system. This
tendency is confirmed by the data for Ar+KCl and Au+Au (figures \ref{fig:ar+kcl_32-74_zmax_delta=3D1/2}
and \ref{fig:au+au_40-80_zmax_delta=3D1}). %
The former has a total mass smaller than both Xe+Sn and Ni+Ni. The
largest fragments produced in central collisions of $^{36}$Ar+KCl
follow quite closely a $\Delta\sim\frac{1}{2}$ scaling behaviour
for beam energies up to at least 52\amev~(Figure \ref{fig:ar+kcl_32-74_zmax_delta=3D1/2}a),
while the data for 74\amev, which deviate from the overall trend,
may indicate a transition to the $\Delta\sim1$ regime occurs somewhere
between these last two available data points. The data for 32--52\amev~exhibit
very similar near-Gaussian probability distributions. Although we
can not have much confidence in a scaling law established for only
two data points let us remark in passing that the scaling behaviour
observed for data at 52 and 74\amev~is approximately a $\Delta\sim1$
scaling with a scaling function of a form similar to that observed
for the other {}``disordered'' regime data, indicating that the
transition energy for this system is probably somewhere between 52
and 74\amev, higher than for the heavier systems.

On the other hand, for the much heavier $^{197}$Au+$^{197}$Au system
a $\Delta=1$ scaling law is observed for fluctuations of the size
of the largest fragment in each event for all studied beam energies
(Figure \ref{fig:au+au_40-80_zmax_delta=3D1}). Even at the lowest
energy (40\amev) the scaling of the entire $Z_{max}$ distribution
with the other energies is excellent. For this system it may be that
the beam energy of the transition from the $\Delta=\frac{1}{2}$ regime,
if it exists, to $\Delta=1$ occurs at a lower energy than the lowest
energy available for study in our data, which may explain why we do
not observe an {}``ordered'' regime for this system. Of course it
is equally possible that no such regime exists for this system. Indeed,
following the interpretation of Sec.\ref{sub:scaling-with-energy}
in terms of the disappearance of fusion-evaporation residues, we would
not expect fusion to occur for a system as heavy as Au+Au at any incident
energy.

\begin{figure}[htbp]
\includegraphics[%
width = \columnwidth]{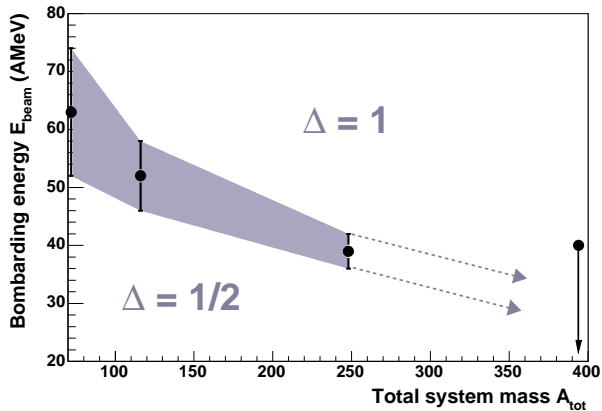}

\caption{Dependence on bombarding energy and total system mass of the frontier
between the two $\Delta$-scaling regimes observed in this work for
very central collisions.\label{fig:transition_energy_vs_mass}}
\end{figure}

We therefore observe that the bombarding energy (or available energy)
at which there is a transition from the {}``ordered'' to the {}``disordered''
regime decreases with increasing total mass of the system under study.
This is represented in Figure \ref{fig:transition_energy_vs_mass}.
If one interprets our results in the framework of the universal fluctuations
theory, i.e. in terms of a 2nd order phase transition, one would expect
the energy at which one phase is replaced by the other to be related
to the critical temperature of the corresponding system. The definition
of this quantity for finite, charged, systems such as nuclei has received
much theoretical attention. On the one hand, in the absence of Coulomb
forces, a pseudo-critical temperature (corresponding to a large peak
in the finite system specific heat) may be defined which is smaller
than the infinite matter value $T_{C}$, and increases as the size
of the system approaches the infinite matter limit \cite{Jaq84:tcrit_finite}.
On the other hand, calculations including the Coulomb repulsion show
that the maximum temperature that an equilibrated hot nucleus can
support , $T_{lim}$, decreases for heavier nuclei, due to their increasing
nuclear charge \cite{Levit:T_limite}. A recent systematic study of
a wide range of data on so-called caloric curves by Natowitz et al
\cite{Natowitz:2001cq} has shown that the temperature and excitation
energy at which a plateau is observed in these curves decreases with
increasing mass of the primary excited nucleus, and that such behaviour
is consistent with theoretical predictions for the limiting temperatures,
$T_{lim}$.

In the present analysis we chose not to study this question in more
detail, as to do so would require: (i) to show that the fragment production
is thermally-driven; (ii) to identify the thermal (sub-)system in
each case; and (iii) to deduce the mass, charge, excitation energy
and temperature of this (sub-)system. Each step would require us to
make important hypotheses about the formation and decay properties
of fragments in the collisions under study, taking us far from our
initial goal of determining as much information as possible on the
nature of fragment production with a minimum number of suppositions.
Moreover, the observed effect, which depends on the entrance channel
total mass and available/bombarding energy may have a completely different
origin. Therefore we will limit ourselves to the observation that
the transition from {}``ordered'' to {}``disordered'' regime takes
place at a lower available energy for systems of greater total mass.

\subsection{Detailed study of the form of the scaling functions\label{sub:form}}

\begin{table*}[htbp]

\caption{Fluctuation scaling exponent $\Delta$, coefficient of skewness $\gamma$,
kurtosis $\kappa$, $\chi^{2}$ for global fits to data with Gaussian
($\Delta\sim\frac{1}{2}$) and Gumbel ($\Delta\sim1$) distributions,
and scaling function tail exponent $\hat{\nu}$($\chi^{2}$) (see
Section \ref{sec:Universal-fluctuations}), for the scaling functions
shown in Figures \ref{fig:xe+sn_zmax_delta=3D1/2_delta=3D1}, \ref{fig:ni+ni_zmax_delta=3D1/2_delta=3D1},
\ref{fig:ar+kcl_32-74_zmax_delta=3D1/2} and \ref{fig:au+au_40-80_zmax_delta=3D1}.\label{tab:scaling_function_parameters}}

\begin{ruledtabular} \begin{tabular}{cc|rrrrrl} System & $E_{inc}$ (AMeV) & $\Delta$ & $\gamma$ & $\kappa$ & $\chi^2$ & $\hat{\nu}$($\chi^2$)\\ \hline  Xe+Sn  & 25--39  & 0.48 & 0.45 & 0.66 & 27 & --\\  Ni+Ni  & 32--52  & 0.61 & 0.48 & 0.64 & 72 & --\\  Ar+KCl & 32--52+  & 0.46 & 0.33 & 0.64 & 63 & --\\ \hline  Au+Au  & 40--80  & 1.00 & 0.83 & 1.39 &  1.8 & 0.92$\pm$.09(1.3)\\  Xe+Sn  & 39--100 & 0.89 & 0.85 & 1.20 & 11 & 1.20$\pm$.03(1.2)\\  Ni+Ni  & 52--90  & 0.88 & 0.84 & 1.30 & 45 & 1.54$\pm$.04(7)\\ \end{tabular} \end{ruledtabular} 
\end{table*}

The results of the analysis for all the systems presented here are
summarized in Table \ref{tab:scaling_function_parameters}. Systems
have been grouped according to their observed $\Delta$-scaling behaviour,
$\Delta\sim\frac{1}{2}$ at the lowest beam energies and $\Delta\sim1$
at the highest. In order to get some quantitative information on the
form of the scaling function in each case, we calculated the overall
coefficient of skewness, $\gamma$, and the kurtosis, $\kappa$, for
each system in each regime, using the following definitions \cite{Roe:statistics}

\begin{eqnarray}
\gamma & = & \frac{<z_{(\Delta)}^{3}>}{\sigma^{3}}\label{eq:skew}\end{eqnarray}
\begin{eqnarray}
\kappa & = & \frac{<z_{(\Delta)}^{4}>}{\sigma^{4}}-3\label{eq:kurtosis}\end{eqnarray}
Larger skewness values indicate more asymmetric distributions, while
the kurtosis measures the deviation of the distribution from the Gaussian
form ($\kappa=0$ being a perfect Gaussian). The values shown in the
table confirm our observation that the higher-energy, $\Delta\sim1$-scaling
data present more asymmetric, less Gaussian probability distributions
than those at lower energy. It can also be seen that, quantitatively,
the scaling functions corresponding to the two regimes for systems
of different masses have very similar forms, confirming the fact that
data for different systems do indeed collapse on to a single distribution.

In the {}``ordered'' regime the fluctuations of the size of the
largest fragment show a significant deviation from a Gaussian distribution,
on the contrary to the near-perfect Gaussian distributions that we
observe for total and fragment multiplicities (Figures \ref{fig:xe+sn_25-100_mtot_delta=3D1/2},
\ref{fig:xe+sn_25-100_mimf_delta=3D1/2} and \ref{fig:ni+ni_32-90_mtot_mimf_delta=3D1/2}).
For the {}``disordered'' regime the deviation from the Gaussian
form is large and the shape of the distribution function is well reproduced
by a Gumbel distribution,\begin{equation}
\Phi(z_{\Delta})\sim\exp-\left(z-\mathrm{e}^{-z}\right)\label{eq:gumbel}\end{equation}

This is shown by the dashed curves in Figures \ref{fig:xe+sn_zmax_delta=3D1/2_delta=3D1}b
and \ref{fig:ni+ni_zmax_delta=3D1/2_delta=3D1}b, and the grey curve
in Figure \ref{fig:au+au_40-80_zmax_delta=3D1}b, which represent
the best fits to these data using Equation \ref{eq:gumbel}. The overall
agreement can be seen to be quite good, and excellent for the Au+Au
data where $\chi^{2}=1.8$ is achieved. Some significant deviations
can be seen however in the tail of the distributions for the lighter
systems Ni+Ni and Xe+Sn: the data seem to have a faster-than-exponential
fall-off for large (positive) fluctuations about the mean value. 

To confirm this in a quantitative manner we performed fits to the
tails of these functions with the asymptotic scaling function form
$\exp-z^{\hat{\nu}}$ (see Table \ref{tab:scaling_function_parameters}).
They show that the exponent $\hat{\nu}$ is very close to 1 for the
Au+Au data, confirming the observation of an exponential tail compatible
with the Gumbel distribution, while for Ni+Ni and Xe+Sn data we find
$1<\hat{\nu}<2$. Let us remark in passing that, just as $\Delta$
seems to increase toward the asymptotic value of 1 with increasing
system mass in the {}``disordered'' regime, it is possible that
$\hat{\nu}$ also has a systematic mass dependence, and decreases
towards an asymptotic value of 1 (exponential tail) for the heaviest
system.

We have clearly and quantitatively established the form of the scaling
functions in the two regimes, and in the data presented here we do
not observe any deviation from these {}``canonical'' forms, quasi-Gaussian
at low energy and quasi-Gumbel at high energy. We do not, therefore,
have any information on the nature of the transition between the two
regimes: as we discussed in Section \ref{sec:Universal-fluctuations}
we cannot be certain of the form of the scaling function at the critical
point, but one may see a sharp decrease of the large-$z$ tails (faster
than $\exp-z^{2}$), or an order parameter distribution (OPD) with
an exponential large-$z$ tail but significantly different from the
Gumbel distribution around the maximum and below \cite{Botet:2004aa}.
On the other hand, for a first-order phase transition with a passage
through the coexistence region, although the scaling behaviour of
the OPD is not well established \cite{Botet:2000dk,Carmona:2001fx},
by definition in this case the order parameter should present a bimodal
(double-humped or very wide) distribution. More detailed study of
existing and new data around the transition energy of 39\amev~(for
the Xe+Sn system) may yet reveal such features.

\subsection{Dependence of the observed scaling behaviour on the violence/centrality
of collisions\label{sub:selection-dependent}}

\begin{figure}[htbp]
\includegraphics[%
width = \columnwidth]{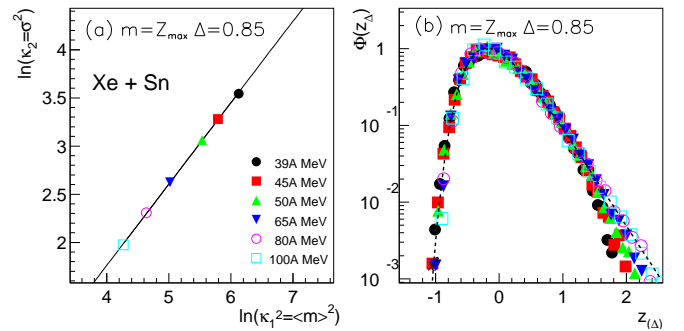}

\caption{(Color online) As Figure \ref{fig:xe+sn_25-100_zmax_delta=3D1},
but for collisions with an estimated centrality of $b<0.2b_{max}$.\label{fig:xe+sn_39-100_zmax_b<0.2bmax}}
\end{figure}

The results presented above come from a wide-ranging set of data concerning
different system energies and masses, but they represent only 1\%
of the total measured cross-section due to our centrality cut. The
representativity of the observed scaling behaviour is far from certain
in this case, as it is well known that data selected by cuts in distribution
tails are prone to serious autocorrelation effects due to conservation
laws \cite{Phair92:ip_filters,Llope95:autocorrelations_imf_mult}.
We therefore have to study the dependence of our findings on the strictness
of our centrality cuts. In Figure \ref{fig:xe+sn_39-100_zmax_b<0.2bmax}
the data for the same Xe+Sn collisions as in Figure \ref{fig:xe+sn_zmax_delta=3D1/2_delta=3D1}b
have been analyzed with a slightly relaxed centrality cut, $b<0.2b_{max}$.
The effect on the apparent value of $\Delta$ can be seen from the
cumulant moments plot, Figure \ref{fig:xe+sn_39-100_zmax_b<0.2bmax}a:
relaxing the centrality condition gives an apparent scaling law with
a smaller value of $\Delta$, which is here $\Delta\approx0.85$ instead
of $\Delta=0.89$ (see Table \ref{tab:scaling_function_parameters}).
A similar dependence on the estimated centrality of collisions was
observed in \cite{Botet:2001qp}, and interpreted in terms of a smaller
excitation energy of fragmenting systems in less central collisions.
However, as Figure \ref{fig:xe+sn_39-100_zmax_b<0.2bmax}b shows,
the $\Delta$-scaling actually begins to break down for these less
central data, as can be seen in the high-$z_{(\Delta)}$ tail of the
{}``scaling'' function.

\begin{figure}[htbp]
\includegraphics[%
width = \columnwidth]{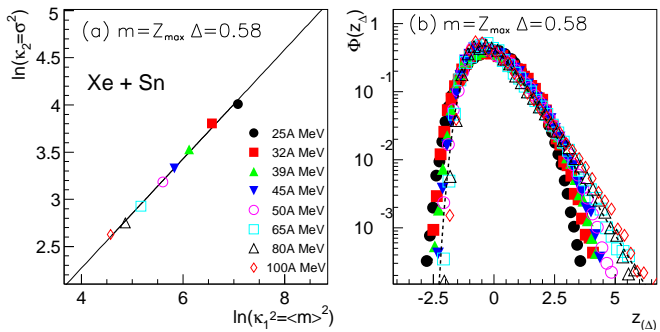}

\caption{(Color online) As Figure \ref{fig:xe+sn_39-100_zmax_b<0.2bmax},
but for collisions with an estimated centrality of $b<0.4b_{max}$.\label{fig:xe+sn_25-100_zmax_b<0.4bmax}}
\end{figure}

If the centrality condition is relaxed even further, $b<0.4b_{max}$
(Figure \ref{fig:xe+sn_25-100_zmax_b<0.4bmax}), we first observe
that now apparently all of the data for Xe+Sn is compatible with an
approximate second-scaling law ($\Delta=0.58$), without any change
of scaling regime between 25 and 100\amev. However we can not assign
the data to a single {}``ordered'' regime because in fact this is
not true $\Delta$-scaling: the different distributions no longer
collapse to a universal curve (Figure \ref{fig:xe+sn_25-100_zmax_b<0.4bmax}b). 

Therefore we observe universal fluctuations and behaviour of the $Z_{max}$
observable compatible with it being an order parameter for nuclear
multifragmentation only for very central collisions of symmetric systems
for which one may suppose a near-to-total overlap of the projectile
and target in the entrance channel. We have not, up to now, observed
an equivalent scaling for quasi-projectile residues in mid-peripheral
to peripheral collisions. This may be because in such reactions the
fragment production is far more sensitive to entrance channel effects,
which vary greatly as a function of the colliding nuclei and their
energies. If so, a meaningful comparison between different systems
is harder to achieve, at least with our rather {}``global'' approach.
In head-on collisions, on the other hand, the fragmentation of the
projectile and target may be virtually a statistical (although not
necessarily thermal) process, far less sensitive to the details of
the reaction, and therefore more amenable to reveal features which
are independent of the system under study.

\section{Summary}

We have studied nuclear multifragmentation data obtained with the
INDRA 4$\pi$ array for collisions of symmetric systems of total mass
$A_{tot}\sim75-400$ at bombarding energies from 25 to 100\amev.
Using the total transverse energy of light charged particles, $E_{t12}$,
as a measure of collision violence, we deduced the evolution of these
reactions with beam energy, impact parameter and system size from
the experimentally measured correlations between the charge of the
largest fragment detected in each event, $Z_{max}$, and $E_{t12}$.
For all data presented in this work these correlations are dominated
by reactions leading to a projectile-like fragment whose size decreases
with increasing collision violence. In the case of the heaviest projectiles
(Xe, Au) fission of the moderately excited quasi-projectile modify
this picture. For higher excitation energies the opening of the quasi-projectile
multifragmentation channel may exhibit a bimodal behaviour which has
been evidenced \cite{Tamain:bimodal} in selecting complete detection
events. On the other hand, in central collisions for a few percent
of the measured cross-section {}``heavy'' residues are produced
in the systems Ar+KCl, Ni+Ni and Xe+Sn, suggesting incomplete fusion
of projectile and target, for beam energies which are not too high
($\leq52$\amev~for Ar+KCl, $\leq40$\amev~for Ni+Ni, $\leq32$\amev~for
Xe+Sn).

Using a model-independent analysis based on the theory of universal
fluctuations of the order parameter for finite systems, we tested
the most violent collisions for signals that the fragment production
may be related to a phase transition. Following the results of \cite{Botet:2001qp}
we first confirmed that, of the two possible order parameters for
a critical fragmentation process, it is the charge (size) of the largest
fragment, $Z_{max}$, and not the total or IMF multiplicities $M_{tot}$
or $M_{Z\geq3}$, which has a behaviour of the scaling properties
of its fluctuations compatible with its being an order parameter for
a critical fragmentation process. Indeed, we have shown that the event-by-event
distribution of $Z_{max}$ allows to sort data into two fluctuation-scaling
regimes defined by the value of the scaling exponent $\Delta$, which
is approximately equal to $\frac{1}{2}$ at low energies and tends
towards the asymptotic value 1 at high energies with increasing total
system size. These regimes are equally well characterised by a distinctive
form of the scaling function $\Phi(z_{(\Delta)})$. At low energies
this function, although more symmetric than that seen at higher energy,
is significantly different from the Gaussian form. The deviation is
quantitatively the same for the three different-sized systems (Ar+KCl,
Ni+Ni and Xe+Sn) for which we observed this low-energy regime. In
the high-energy regime the scaling function is more asymmetric and
tends towards the asymptotic form of the Gumbel distribution with
increasing system mass. This evolution concerns mainly the large-$Z_{max}$
tail of the distribution which falls off more slowly for heavier systems,
becoming exponential. For the Au+Au system $\Phi(z_{(\Delta)})$ is
an almost perfect Gumbel distribution.

The bombarding energy at which the passage from one regime to the
other is situated decreases as the total system mass (and charge)
increases. This is the contrary of the expected behaviour if this
transition were related to the critical temperature of the systems
under study, which would increase with the mass, but is on the other
hand consistent with the observation of decreasing limiting temperatures
for finite nuclei \cite{Natowitz:2001cq} due to the increase in Coulomb
energy for heavy nuclei. However, we can only speculate whether this
is the reason for our observation, as our analysis does not depend
on any model of fragment production such as supposing it to be equilibrated
or thermally-driven, and therefore it does not give any information
on whether these conditions are met or not. On the other hand, data
strongly suggest that the dependence on entrance channel mass and
bombarding energy of the two regimes is closely linked to the disappearance
of heavy residues in central collisions.

Although this analysis allows to establish the existence of two distinct
regimes in multifragmentation reactions based on the scaling properties
of the fluctuations of the largest fragment, it has not given any
information on the passage from one regime to the other. In the data
studied so far we observe only order parameter distributions which
are compatible with weakly-correlated systems, i.e. far from the critical
point. It may be that the general survey of a wide range of data presented
here is not sufficiently detailed to reveal such features, or that
data taken in small bombarding energy steps around the {}``transition
energy'' is necessary to track the evolution of the order parameter
distribution. Such additional data for the Xe+Sn system have been
measured recently and are currently under analysis. It is our feeling
that the present work constitutes a solid basis for further study
of the question of criticality in nuclear multifragmentation data.

\begin{acknowledgments}
We would like to thank M. P\l oszajczak for fruitful discussions
and the support and interest he has shown in our work.

We thank also the staff of the GANIL and GSI Accelerator facilities
for their support during the experiments. This work was supported
by the European Community under contract No. ERBFMGECT950083, and
by the Commissariat à l'Energie Atomique, the Centre National de la
Recherche Scientifique, the Ministère de l'Education Nationale, the
Conseil Régional de Basse Normandie and the Association Française
des Femmes Diplômées des Universités.
\end{acknowledgments}

\bibliographystyle{revtex}

\end{document}